\documentclass[usenatbib]{mn2e}

\usepackage{times}
\usepackage{graphicx}
\usepackage{amssymb}

\begin{document}

\title{P/2006 VW$_{139}$: A Main-Belt Comet Born in an Asteroid Collision?}

\author[Bojan\ Novakovi\'c et al.]{Bojan~Novakovi\'c,$^1$\thanks{E-mail: bojan@matf.bg.ac.rs} 
Henry H.\ Hsieh$^2$\thanks{Hubble Fellow} and 
Alberto Cellino$^3$\\
$^1$ Department of Astronomy, Faculty of Mathematics, University of Belgrade,
Studentski trg 16, 11000 Belgrade, Serbia \\
$^2$ Institute for Astronomy, University of Hawaii, 2680 Woodlawn
Drive, Honolulu, HI 96822, USA \\
$^3$ INAF--Osservatorio Astronomico di Torino, Via Osservatorio 20, I-10025
Pino Torinese, Italy}


\pagerange{\pageref{firstpage}--\pageref{lastpage}} \pubyear{2012}

\maketitle 

\label{firstpage}

\begin{abstract}
In this paper we apply different methods to examine the possibility that a small group of 24 asteroids
dynamically linked to main-belt comet P/2006 VW$_{139}$, recently discovered by the Pan-STARRS1 survey
telescope, shares a common physical origin. By applying the Hierarchical Clustering and 
Backward Integration methods, we find strong evidence that 11 of these asteroids form a sub-group
which likely originated in a recent collision event, and that this group includes P/2006 VW$_{139}$.
The objects not found to be part of the 11-member sub-group, which we designate as the P/2006 VW$_{139}$ family, 
were either found to be dynamically unstable, or these are likely interlopers which should be expected due to the 
close proximity of the Themis family. As we demonstrated, statistical significance of P/2006 VW$_{139}$ family
is $> 99\%$.
We determine the age of the family to be $7.5\pm0.3$~Myr, and estimate the diameter of the parent body to be $\sim11$~km.
Results show that the family is produced by an impact
which can be best characterized as a transition from catastrophic to cratering regime.
The dynamical environment of this family is studied as well, including the identification of 
the most influential mean motion and secular resonances in the region. 
Our findings make P/2006 VW$_{139}$ now the second main-belt comet to be 
dynamically associated with a young asteroid family, a fact with important implications
for the origin and activation mechanism of such objects.
\end{abstract}



\section{Introduction}

Main belt comets (MBCs) are objects that are dynamically indistinguishable from main belt asteroids,
but which exhibit comet-like activity due to the sublimation of volatile ice
\citep{Hsieh2006,bertini2011}. 
To date, six such objects have been discovered \citep{Hsieh2012b}. MBCs are important because they represent a 
new reservoir of comets in the Solar system, and may give insights into the role of main-belt objects
in the primordial delivery of water to the Earth as well as provide constraints on the 
composition of the protosolar disk.

Dynamical simulations performed by \citet{Fernandez2002} show that it is unlikely that present-day objects with
main-belt orbits originated in the Kuiper Belt, assuming the current configuration of the major planets.
This indicates that MBCs are likely native to the main asteroid belt. 
Still, \citet{Levison2009} show that during the early violent dynamical evolution of 
giant planet orbits, as required by the so-called Nice model \citep{menios,morby,gomes}, some icy 
trans-Neptunian bodies may have been implanted in the asteroid belt. It is expected that most of these objects are located 
in the outer belt, but they might be found anywhere with semi-major axes larger than about 2.6~au.

Numerical simulations performed to assess the dynamical stability of MBCs suggest that the orbits of 
the currently known objects are stable, indicating that they are likely native to their current locations 
\citep{haghighipour2009,Hsieh2012a,Hsieh2012b}. The only exception is P/2008 R1 (Garradd), which appears to be 
dynamically unstable \citep{jewitt2009}. These simulations did not take into
account non-gravitational effects (e.g., the Yarkovsky effect), however, that may play an important role
in the long-term stability of the MBCs.

Wherever MBCs originate from, their activity, although still not reliably understood, is likely triggered 
by the impact-excavation of subsurface ice \citep{Hsieh2004,jewitt2012} because completely exposed surface ice is unstable
against sublimation at their heliocentric distances over Gyr time-scales.  According to \citet{Hsieh2009} and \citet{capria2012},
this hypothesis is in reasonable agreement with the present-day impact frequency in the main belt. 

Thermal modelling by \citet{Schorghofer2008} and \citet{Prialnik2009} shows that buried ice on 
a main-belt comet can in fact survive over the 
life of the solar system.  However, while thermal devolatization may not preclude the existence of present-day 
ice (and therefore MBCs) in the main asteroid belt, there is the 
additional problem of collisional devolatization.  Each time an impact 
triggers activity in a MBC by exposing a small amount of subsurface ice to 
direct solar heating, that particular area of the surface is effectively 
devolatilized once that ice has sublimated away. Using 
estimates of the active areas required to produce the activity observed 
for MBCs, Hsieh (2009) determined that the observed activity was most 
likely triggered by approximately metre-sized impactors. Using work by 
\citet{cheng2004} and \citet{bottke2005b}, it was then estimated that impacts 
by objects of this size in the main belt should occur roughly every 
$10^4$~yrs. Over Gyr time-scales, the cumulative effect of such impacts 
could be to devolatilize a significant portion of the surface of an 
ice-bearing asteroid.  More deeply buried ice could persist, safe from 
both thermal and collisional depletion, but would also therefore be 
inaccessible by activity-triggering impacts. Such asteroids would then be 
just as unlikely to exhibit activity in the present day and be identified 
as MBCs as other non-ice-bearing asteroids.

Because of these reasons, it is important to understand the dynamical 
environment of each MBC and any possible links to collisionally-formed 
asteroid families. Asteroid families are believed to originate in the 
catastrophic disruption of large asteroids \citep[e.g.][]{Zappala2002}, 
and are a fascinating and challenging subject in themselves. A few tens 
of families have been discovered to date across the main belt \citep[e.g.][]{Zappala1995,MD2005}, 
providing insights into the collisional history of 
the main belt \citep[e.g.,][]{Marzari1999,bottke2005}, disruption 
events over a size range inaccessible to laboratory experiments 
\citep[e.g.,][]{michel2003,durda2007}, the mineralogical structure of 
their parent bodies \citep[e.g.,][]{cellino2002}, and many other subjects.

Previous analyses show that three of the six currently known MBCs 
(133P/Elst-Pizarro, 176P/LINEAR, and P/2006 VW$_{139}$) are associated 
with the large and old Themis family \citep{Hsieh2006,Hsieh2012b}, 
which is thought to have formed from the catastrophic disruption 
of a $\sim$400~km parent body $2.5\pm1.0$~Gyr ago \citep{Nesvorny2003}.  
Additionally, a fourth MBC (238P/Read) is close to the Themis 
family in orbital element phase space, and may have once belonged to the 
family \citep{haghighipour2009}. More significantly, 133P has been found to 
additionally belong to the small and young Beagle family, which is thought 
to have formed less than 10~Myr ago \citep{Nesvorny2008}. This finding is significant because,
using the devolatization rate derived by \citet{Hsieh2009},
if one assumes 133P is a primordial member of the Themis family, placing 
its age at $\sim$2.5~Gyr, less than 10 per cent of 133P's surface would be expected to be 
unaffected by impacts by the present day.  In contrast, 99 per cent of 
133P's surface would be expected to remain unimpacted if the object is 
assumed to only be 10~Myr old (i.e., the age of the Beagle family).  This 
possible link to young asteroid families is supported by the finding by 
\citet{Hsieh2012a} that P/2010 R2 (La Sagra) belongs to a small cluster 
that may have a common collisional origin.

Comet-like activity in main-belt asteroid (300163) 2006 VW$_{139}$ was discovered 
by the Pan-STARRS1 survey telescope on Haleakala in Hawaii on 2011 November 5 \citep{Hsieh2011},
with preliminary analysis of its activity \citep{Hsieh2012b} showing that it is likely to be a genuine
comet (i.e., exhibiting activity due to sublimation), rather than a disrupted asteroid
(i.e., whose activity is due to a recent impact) such as P/2010 A2 (LINEAR)
\citep[e.g.,][]{Jewitt2010,Snodgrass2010} and (596) Scheila \citep{Jewitt2011,Bodewits2011}.
\citep{Hsieh2012b} also examined P/2006 VW$_{139}$'s possible links to dynamical asteroid families
and found that it is dynamically associated with the Themis family, but
that it may also belong to a smaller sub-group that could represent a new
previously-unidentified young asteroid family.

In this paper, we conduct a detailed investigation into whether objects dynamically
linked with P/2006 VW$_{139}$ in fact have a common physical origin. First, we identify members of the
group using the Hierarchical Clustering Method (HCM). Then we use
the Backward Integration Method \citep[BIM;][]{Nesvorny2002} in order to further test
the reliability of the group and to estimate its age.  Finally, we analyse the characteristics
of this newly discovered family in greater detail.

\section{Linking P/2006 VW$_{139}$ to dynamical families}

We start our investigation by identifying the objects that are dynamically 
associated with P/2006 VW$_{139}$. The identification of asteroid families is usually done in
proper orbital element space using proper semi-major axes ($a_{p}$), proper eccentricities ($e_{p}$)
and sines of proper inclination ($\sin(i_{p})$).
For this analysis, we applied the HCM \citep{Zappala1990,Zappala1994}
to analytically-determined 
proper orbital elements \citep{MilKne1990,MilKne1994} available at the \textit{AstDys} web 
page\footnote{http://hamilton.dm.unipi.it/astdys/index.php?pc=5}, where we obtained 
elements for 398,841 asteroids (as of January 2012).

\begin{figure}
\includegraphics[angle=-90,scale=.30]{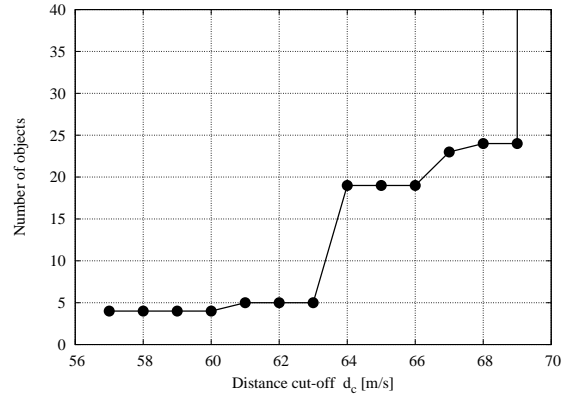}
\caption{Number of asteroids associated with P/2006 VW$_{139}$ as a
function of cut-off distance (in velocity space), expressed in m~s$^{-1}$.
Two critical cut-off distance values are (i) 64~m~s$^{-1}$ when the number of dynamically associated objects increases
significantly, and (ii) 70~m~s$^{-1}$ when the group around P/2006 VW$_{139}$ merges
with the Themis family.
A potential sub-family within the Themis family is reasonably well defined in between these two critical values.
}\label{f:nfv}
\end{figure}

The results obtained by the HCM are the same as those obtained by \citet{Hsieh2012b}, although here we
use an updated proper element catalogue.
Fig.~\ref{f:nfv} shows the number of asteroids dynamically associated with
P/2006 VW$_{139}$ as a function of the cut-off distance, $d_{c}$ (in velocity space), the standard metric in the HCM.
An intriguing sub-grouping appears at $d_c>63$~m~s$^{-1}$ before then merging with the Themis family at $d_{c} = 70$~m~s$^{-1}$.
A complete list of asteroids that belong to this group is given in Table~\ref{t:300163list}. Assuming that a cut-off value of $d_c=70$~m~s$^{-1}$
used by \citet{nes2010PDS} to characterize the Themis family is still appropriate, the sub-group around
P/2006 VW$_{139}$ is formally a part of this family. However, the latter result should be interpreted with some 
caution because the group occupies the outskirts of the Themis family. Also, one should bear in mind that the
cut-off value appropriate for characterizing an asteroid family tends to decrease as the number of known
asteroids increases because the density of background objects becomes higher. Consequently, the appropriate value
of $d_{c}$ for the Themis family may be slightly smaller today than at the time of \citet{nes2010PDS}
work. This is supported by the fact that list of the Themis family members
produced by \citet{nes2010PDS} does not include the asteroids from P/2006 VW$_{139}$ group,
although most of these objects were known at that time.

\begin{table*}
\begin{minipage}{144mm}
\centering
\caption{Members of the P/2006 VW$_{139}$ Group}
\label{t:300163list}
\begin{tabular}{lccccccrc}
\hline
Object
 & $a_p$\footnote{Proper semi-major axis, in a.u.} & $e_p$\footnote{Proper eccentricity}
 & $\sin(i_p)$\footnote{Sine of proper inclination}
 & $H$\footnote{Absolute $V$-band magnitude, from the AstDys website}
 & $p_{v}\pm\sigma_{p_{v}}$\footnote{Geometric $V$-band albedo, from \citet{wise} if available; otherwise the mean value for the group is given in brackets}
 & $D$\footnote{Diameter, in km}
 & $T_{\rm lyap}$\footnote{Lyapunov time, in kyr}
 & Clustering\footnote{Notes about nature of clustering of secular angles, with ``Y'' indicating that clustering
    of secular angles exist, percentage values indicating the chance of clustering derived using
    clones, and ``U'' indicating unstable objects} \\
\hline
15156           & 3.05542 & 0.1647 & 0.0437 & 13.4 & $0.1032\pm0.0207$& 8.7 &  151.5   &  Y \\
32868           & 3.06753 & 0.1639 & 0.0467 & 14.1 &      (0.07)      & 7.5 &   45.3   &  U \\
51798           & 3.05414 & 0.1668 & 0.0451 & 14.8 &      (0.07)      & 5.5 &   12.0   &  U \\
60320           & 3.05300 & 0.1602 & 0.0398 & 14.8 &      (0.07)      & 5.5 &   91.7   &  U \\
117310          & 3.04833 & 0.1651 & 0.0443 & 14.7 & $0.0587\pm0.0084$& 6.2 &  3125.0  &  $12\%$ \\
144767          & 3.05948 & 0.1635 & 0.0464 & 15.8 & $0.0770\pm0.0375$& 3.3 &  2325.6  &  Y \\
220751          & 3.04909 & 0.1608 & 0.0433 & 16.6 &      (0.07)      & 2.5 &  1250.0  &  Y\\
221660          & 3.03339 & 0.1645 & 0.0374 & 15.2 &      (0.07)      & 4.5 &   5.4    &  U \\
221760          & 3.04065 & 0.1621 & 0.0407 & 15.8 &      (0.07)      & 3.5 &  3703.7  &  Y \\
250710          & 3.04587 & 0.1612 & 0.0378 & 15.9 &      (0.07)      & 3.4 &  408.2   &  $6\%$ \\
255251          & 3.04918 & 0.1654 & 0.0463 & 17.0 &      (0.07)      & 2.0 &  7142.8  &  Y \\
255333          & 3.05631 & 0.1626 & 0.0489 & 15.9 &      (0.07)      & 3.3 &  970.9   &  $0\%$ \\
257688          & 3.05167 & 0.1629 & 0.0519 & 15.4 &      (0.07)      & 4.2 &  4166.7  &  Y \\
261697          & 3.05010 & 0.1637 & 0.0405 & 16.5 &      (0.07)      & 2.5 &  699.3   &  Y \\
294724          & 3.06029 & 0.1625 & 0.0444 & 16.8 &      (0.07)      & 2.2 &  1428.6  &  Y \\
300163          & 3.05350 & 0.1613 & 0.0379 & 16.2 &      (0.07)      & 2.9 &  2083.3  &  Y \\
305976          & 3.04891 & 0.1653 & 0.0430 & 16.5 &      (0.07)      & 2.6 &  625.0   &  $0\%$ \\
2002 VU$_{137}$ & 3.02924 & 0.1627 & 0.0384 & 16.9 &      (0.07)      & 2.1 &  4.8     &  U \\
2005 TH$_{194}$ & 3.05136 & 0.1611 & 0.0409 & 16.6 & $0.0411\pm0.0165$& 3.2 &  6666.7  &  $0\%$ \\
2005 YC$_{149}$ & 3.05162 & 0.1613 & 0.0510 & 17.5 &      (0.07)      & 1.6 &  4347.8  &  Y \\
2007 DP$_{70}$  & 3.04943 & 0.1620 & 0.0505 & 16.8 &      (0.07)      & 2.2 &  3333.3  &  $16\%$   \\
2007 UC$_{1}$   & 3.04305 & 0.1643 & 0.0418 & 16.6 &      (0.07)      & 2.4 &   5555.6 &  $18\%$   \\
2008 FG$_{14}$  & 3.02182 & 0.1626 & 0.0364 & 17.4 &      (0.07)      & 1.7 &  5555.6  &  Y   \\
2011 UB$_{54}$  & 3.03709 & 0.1625 & 0.0378 & 16.6 &      (0.07)      & 2.4 &  800.0   &  $19\%$   \\
\hline
\end{tabular}
\end{minipage}
\end{table*}

\begin{figure}
\includegraphics[angle=-90,scale=.32]{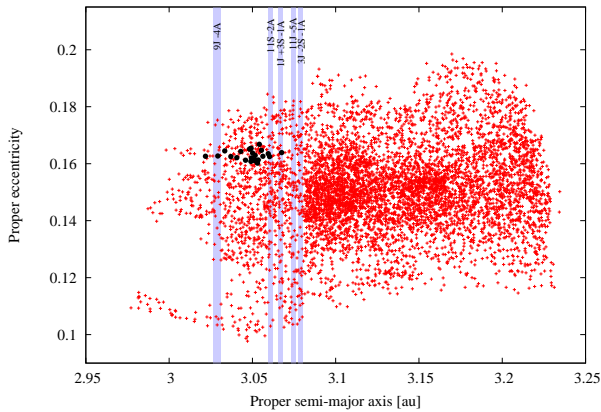}
\caption{The distribution of the members of Themis family (small crosses) 
and the sub-group around P/2006 VW$_{139}$ (filled circles) in the ($a_{p}$,$e_{p}$) 
plane. The most notable two- and three-body mean motion resonances located 
close to P/2006 VW$_{139}$ group are marked as 
vertical dashed areas.}\label{f:ae}
\end{figure}

\begin{figure}
\includegraphics[angle=-90,scale=.32]{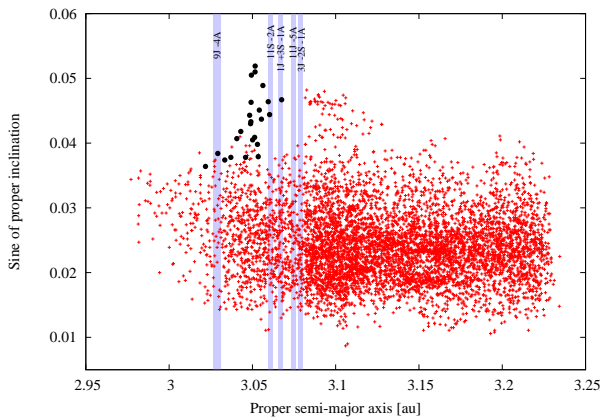}
\caption{The same as in Fig.~\ref{f:ae}, but in the ($a_{p}$,$\sin(i_{p})$) plane.
Note that group around P/2006 VW$_{139}$ is located on the edge of the Themis family.}
\label{f:ai}
\end{figure}

On the other hand, as can be seen in Figs.~\ref{f:ae} and~\ref{f:ai}, the P/2006 VW$_{139}$ group is 
separated from the main part of Themis family by several two- and three-body\footnote{Three-body 
mean motion resonances are those whose resonant angle is defined as a linear 
combination of the mean longitudes of two planets and an asteroid \citep{Nesvorny1998}.} mean motion 
resonances (MMRs) and actually may be a real part of the Themis family simply separated by the MMRs.
 
\begin{figure}
\includegraphics[angle=0,scale=.38]{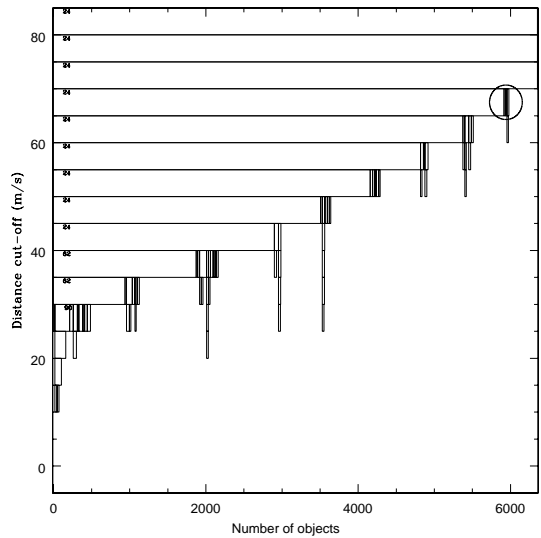}
\caption{The stalactite diagram for 6374 Themis family members. 
Only the groups with at least $N_{crit}$ = 19 members are shown. Note that P/2006 VW$_{139}$ group
is visible only in the one step, at 65~m~s$^{-1}$ (its location is emphasised by the circle).}
\label{f:stalactite}
\end{figure}

To visualize the structure of the whole Themis family,
we generate a stalactite diagram for the population of 6374 family members (Fig.~\ref{f:stalactite}). This type of diagram
is extensively used in other family identification papers \citep[e.g.,][]{Zappala1990,Zappala1994,Zappala1995,Novakovic2011},
and is a highly effective means of visually displaying the structure of a given region.
Here, however, we have an opposite example: a group, which we show later is likely real,
is visible only in the one step used in the diagram, and would not be considered a family
based on standard criteria for family identification.
This situation is a good illustration of the limitation imposed to this approach in the region
occupied by large and dense asteroid families such as Themis.

The density of background objects in this region is high, and it is exceptionally 
difficult to separate nearby families due to the phenomenon of chaining.\footnote{A well-known 
drawback of the HCM based on the single linkage rule is the so-called 'chaining' phenomenon: 
first concentrations naturally 
tend to incorporate nearby groups, thus forming a 'chain' \citep{Zappala1994,Everitt2001}.}
Because of that, we are unfortunately unable to draw conclusive result whether this group belongs
to the Themis family or not. However, this problem does not affect our further investigation of the 
P/2006 VW$_{139}$ group.

\section{The P/2006 VW$_{139}$ family}

\subsection{Membership and age from backward integrations}
\label{ss:bim}

To obtain a preliminary sense of the dynamical stability of
asteroids that belong to the group, we calculate their Lyapunov times ($T_{\rm lyap}$) following 
the methodology described in \citet{MilNob92}. Our results (see Table~\ref{t:300163list}) 
show that the orbits of 19 objects can be considered stable ($T_{\rm lyap}>100$ kyr).
Two unstable objects (221660 and 2002 VU$_{137}$) are likely trapped inside the 9J:4A resonance,
one (51798) probably interacts with the 20J:9A resonance, and another (32868) appears to
be destabilized by the 1J+3S$-$1A three-body MMR. The asteroid 60320 is located near, but outside 
the 20J:9A resonance (see Figs.~\ref{f:ae2} and \ref{f:ai2}). Its instability may be caused by an unidentified MMR
of high order, or it may interact weakly with the 20J:9A resonance. We also note though that Lyapunov time of this
asteroid is significantly longer than in the case of other unstable objects.
More will be said about dynamical stability in Section~\ref{ss:char}.

The fact that 19 out of 24 objects associated to the group around P/2006 VW$_{139}$ are 
stable makes it possible to apply the BIM to search for possible clustering 
of secular angles in the past. 
As in the analysis of the P/La Sagra cluster presented by \citet{Hsieh2012a}, the purpose of  
the BIM analysis here is two-fold:
if suspicious clusterings in the longitude of the ascending node
($\Omega$) and the argument of perihelion ($\omega$) can be found, this
analysis would not only allow us to estimate the age of
the group, but would also provide support for the hypothesis that these objects 
share a common physical origin.

As mentioned above, due to the proximity of the group around P/2006 VW$_{139}$ 
to the Themis family, a distinction between the members of these two groups is very difficult to discern.
Therefore, the probability of linking random interlopers with the P/2006 VW$_{139}$ 
group is much higher than in the case of ``typical'' families,
where the fraction of interlopers is usually below 10 per cent \citep{Migliorini1995}.
In some cases though, the fraction of interlopers may be as high as $\sim$25 per cent,
as in the case of the Eucharis asteroid family \citep{Novakovic2011}. 

As the members of the group cannot be reliably identified a priori,
application of the BIM is complicated because the presence of
interlopers could mask possible clustering that would detected if
only real members were considered. Given that, we take a slightly unusual approach,
where instead of searching for clustering that involves all asteroids linked
to the group, we search for clusterings using sub-samples
of potential family members. This method allows us
not only to estimate the age of the group, but
also to characterize the family by distinguishing individual members
from interlopers.\footnote{A similar idea was proposed by
\citet{Marcus2011} to search for families among Kuiper belt objects.} 
We name this technique the Selective Backward Integration Method (SBIM).

Specifically, we perform the following four steps:

(i) Starting with the list of dynamically linked group members, as identified by the HCM using 
the largest cut-off distance ($d_c=69$~m~s$^{-1}$) for which the group remains separated from 
Themis family, unstable asteroids are removed, i.e. those with $T_{\rm lyap}<100$~kyr.
This allows us to search for possible clustering among as many asteroids as possible.

(ii) Orbits of stable bodies are numerically integrated using the public domain 
\textit{ORBIT9} software \citep{MilNob88} embedded in the multi-purpose \textit{OrbFit} package.
The dynamical model is purely gravitational and includes four major planets, from Jupiter to 
Neptune, as perturbing bodies. To account for the indirect effect of the 
inner planets, their masses are added to the mass of the Sun and the 
barycentric correction is applied to the initial conditions \citep{MK1992}. 
Orbits of all objects are propagated over 20~Myr in the past. A online digital filtering
is applied to the output of numerical integrations in order to remove short 
periodic perturbations. In this way, mean orbital elements \citep{Kne2000} are obtained,
which can then used to search for possible clustering of the secular angles.

(iii) For the 19 stable asteroids in the P/2006 VW$_{139}$ group, we check whether a clustering within $40^\circ$
in at least one of the secular angles can be found for any sub-group with at least 
10 objects.  The choice of the limit of $40^\circ$ is based on previous results obtained
in application of the BIM \citep{Nesvorny2002,Tsiganis2007,Novakovic2010a},
while the minimum number of objects is chosen assuming a maximum interloper fraction of 50 per cent.

(iv) For each stable object that does not cluster, a set of orbit and ``yarko''
clones is generated to check whether the orbit uncertainty or the Yarkovsky effect \citep{bottke2006} 
may be the reason why clustering is not observed. In particular, ten orbit clones are
drawn from 3-$\sigma$ interval of the uncertainties of orbital elements listed in Table~\ref{t:osc}, 
assuming Gaussian distributions.  Then, for each orbit clone, 10 yarko clones uniformly distributed
over the interval $\pm (da/dt)_{max}$ are generated, where $(da/dt)_{max}$ is the maximum value of 
the semi-major axis drift speed due to the Yarkovsky force. Using a model of the
Yarkovsky effect \citep{vok1998,vok1999,vokfar1999}, 
and assuming thermal parameters appropriate for $C$-type objects \citep{broz2008}, we estimate the 
maximum possible drift speed in the proper semi-major axis due to the Yarkovsky thermal 
effect, for a body of 1~km in diameter, to be $(da/dt)_{max}=4 \times 10^{-4}$~au/Myr.
In this way, a total of 100 clones of each object are produced.
The orbits of all clones are numerically integrated in the framework of the same model
as nominal orbits, but this time with the Yarkovsky force included in the model.\footnote{For 
simplicity, the Yarkovsky effect is approximated in terms of a pure along-track acceleration, 
producing the same average semi-major axis drift speed $da/dt$ as expected from theory.} Finally, a possible
clustering of the secular angles of each of the clones, and that of the asteroids for which
clustering of secular angles is observed in step (iii) is checked. The probability that the stable asteroids whose
nominal orbits do not cluster should have exhibited clustering is estimated using the fraction of clones that exhibit
clustering. 

Applying the above procedure, among 19 asteroids with $T_{\rm lyap}>100$~kyr, we find a sub-group of  
11 objects that exhibit clustering in both $\Omega$ and $\omega$ about 7.5 Myr in the past.
The clustering of $\Omega$ is the only prominent one within the last 20~Myr and is very deep (within $32^\circ$).
As such, we consider it to be a highly reliable event.

On the other hand, although there are several clusterings of $\omega$,
none is particularly deep. This, however, is unsurprising, since we expect to find 
tighter clusterings of $\Omega$ than of $\omega$. 
The reason is that, in the region we consider, the secular frequency\footnote{The $s$, $g$ and $g-s$ 
are frequencies of longitude of the ascending node ($\Omega$), longitude of perihelion ($\varpi$) and argument of
perihelion ($\omega$) respectively.} gradient $\partial s/ \partial a_{p}$ 
is more than a factor of two smaller than $\partial (g-s)/ \partial a_{p}$.
Thus, small changes in $a$ (e.g., due to the Yarkovsky effect) 
cause much quicker dispersions of $\omega$ in a way that is difficult to follow.
Similar situations were seen in the cases of the Veritas, Beagle and Theobalda families 
\citep{Nesvorny2002,Nesvorny2008,Novakovic2010a},
for which the use of $\omega$ to search for clustering was ultimately impossible.

\begin{figure}
\includegraphics[angle=-90,scale=.32]{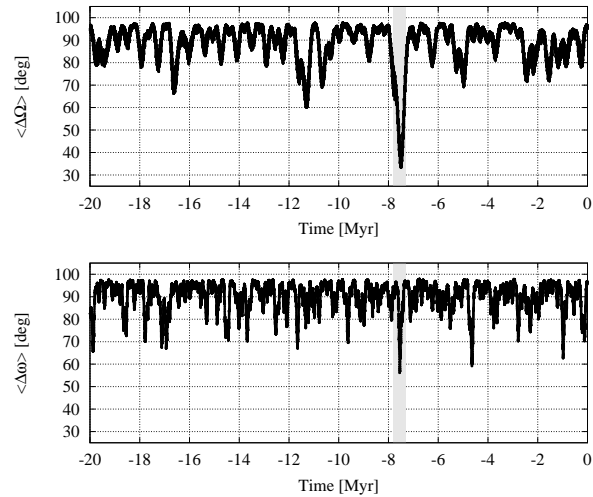}
\caption{The average of differences in the mean longitudes of the ascending nodes $\Omega$ (top),
and arguments of perihelion $\omega$ (bottom), 
for 11 dynamically stable asteroids that belong to the P/2006 VW$_{139}$ group. The results are 
obtained in a purely gravitational model. The most important feature (visible in both angles) 
is clustering at $7.5\pm0.3$~Myr ago, the likely age of the sub-group.}
\label{f:angles}
\end{figure}

Due to the aforementioned reasons, the remarkable match found between the deepest clusterings in
$\Omega$ and $\omega$ is a strong indication that they are not the result of random fluctuations,
but rather an indication that the objects considered share a common origin. We conclude that this
group of 11 asteroids including P/2006 VW$_{139}$ is a real asteroid family formed
$7.5\pm0.3$~Myr ago.\footnote{We use the phrase ``P/2006 VW$_{139}$ group'' to refer to the 
group of 24 asteroids dynamically linked
by the HCM, while ``the P/2006 VW$_{139}$ family'' is used to refer only to those objects whose 
secular angles ($\Omega$ and $\omega$) converge towards the same value.}

\begin{table*}
\begin{minipage}{150mm}
\centering
\caption{Stable members of the P/2006 VW$_{139}$ group whose secular angles ($\Omega$ and $\omega$) do not cluster.}
\label{t:osc}
\begin{tabular}{lcccrrr}
\hline
Object
 & $a$\footnote{Osculating semi-major axis and formal uncertainty, in au}
 & $e$\footnote{Osculating eccentricity and formal uncertainty}
 & $i$\footnote{Osculating inclination and formal uncertainty, in deg}
 & $\Omega$\footnote{Osculating longitude of ascending node, in deg}
 & $\omega$\footnote{Osculating argument of perihelion, in deg}
 & $M$\footnote{Osculating mean anomaly, in deg} \\
 & $\sigma_{a}$ & $\sigma_{e}$ &  $\sigma_{i}$ & $\sigma_{\Omega}$ & $\sigma_{\omega}$ & $\sigma_{M}$ \\
\hline
117310          & 3.0520756061 & 0.1954589262 & 3.67207188 &  89.08283896 & 297.03599694 & 162.03649209 \\
                & 0.0000000861 & 0.0000001940 & 0.00001306 &   0.00024470 &   0.00024860 &   0.00003293 \\
250710          & 3.0406037871 & 0.1972500111 & 1.79346026 & 187.48581675 & 156.96170804 &  90.72340929 \\
                & 0.0000000937 & 0.0000001605 & 0.00001632 &   0.00032530 &   0.00032990 &   0.00004392 \\
255333          & 3.0533298729 & 0.1298456130 & 1.53037270 & 298.13154200 & 273.37048752 & 284.24814758 \\
                & 0.0000001916 & 0.0000002877 & 0.00001458 &   0.00060860 &   0.00062280 &   0.00013090 \\
305976          & 3.0625856084 & 0.1226730090 & 1.61519083 & 215.05214592 & 343.98604328 & 221.68737391 \\
                & 0.0000002950 & 0.0000001935 & 0.00001956 &   0.00054650 &   0.00055560 &   0.00009934 \\
2005 TH$_{194}$ & 3.0455153624 & 0.2008262940 & 1.99615696 & 191.88575916 & 172.34882669 & 110.98328885 \\
                & 0.0000010430 & 0.0000043270 & 0.00003527 &   0.00067090 &   0.00288200 &   0.00169800 \\
2007 DP$_{70}$  & 3.0516613516 & 0.1343786668 & 3.66093739 & 145.64448060 & 346.38036314 &   4.96950838 \\
                & 0.0000003648 & 0.0000005328 & 0.00004180 &   0.00043090 &   0.00055770 &   0.00028460 \\
2007 UC$_{1}$   & 3.0382721562 & 0.1874885082 & 2.62463578 & 165.94090018 & 257.24768301 & 275.35880634 \\
                & 0.0000004401 & 0.0000008379 & 0.00002666 &   0.00036510 &   0.00062120 &   0.00033210 \\
2011 UB$_{54}$  & 3.0330485525 & 0.2025324904 & 2.51664219 &  36.42917884 & 332.62952742 &  52.94143685 \\
                & 0.0000000996 & 0.0000012870 & 0.00005707 &   0.00059980 &   0.00071090 &   0.00022170 \\
\hline
\end{tabular}
\end{minipage}
\end{table*}

The last step in our analysis of clusterings of secular angles is to check whether
orbit uncertainty or the Yarkovsky effect could be the reason why several stable
asteroids linked to the group by the HCM do not cluster. This is done following
the procedure described above in step (iv). The results of this analysis, presented in the 
last column of Table~\ref{t:300163list}, show that this is unlikely for any of the objects,
with probabilities below 20 per cent even in the most promising case of asteroid 2011 UB$_{54}$. 
Thus, only 11 asteroids can be reliably associated with the P/2006 VW$_{139}$ family at this time. 

\subsection{Statistical significance of the new family}
\label{ss:stat}

An important step in our analysis is to estimate a level of statistical significance of 
the P/2006 VW$_{139}$ family. Because our conclusion that family is indeed real is mostly based on
our findings of the clusterings of secular angles, we tested an hypothesis that these clusterings, 
simultaneously in both angles, can occur by a chance. 

In particular the following test was performed. In the same region as occupied by the P/2006 VW$_{139}$ 
group, in terms of osculating orbital elements (3.025 $< a <$ 3.075 au, 0.12 $< e <$ 0.21 and 1.0 $< i <$ 4.5$^\circ$), 
we generated 2500 uniformly distributed test particles. For these particles we followed more
or less the same approach as for the members of the P/2006 VW$_{139}$ group.
Their orbits were integrated for 10 Myr and subsequently the Lyapunov times were determined. 
Then, 1900 test particles on stable orbits were selected and divided in 100 sets (each containing 19 particles). 
For these sets we checked whether clusterings within $32^\circ$ and 55$^{\circ}$, in 
$\Omega$ and $\omega$ respectively, can be found for any smaller subset with 11 objects.
The characteristics of requested clusterings as well as the number of objects are the same ones as in 
the case of the P/2006 VW$_{139}$ family. Also, the fact that test particles are located in the same region
of the main belt as the real family members, means that the dynamical conditions are the same, e.g. differential 
precessions of their secular angles are very similar. 

Performing these tests we did not find any appropriate clustering among 100 groups formed by the test particles.
A clustering in $\omega$, within 55$^{\circ}$, occurs at least once per 10 Myr in about 50 per cent of the cases.
However, a clustering in $\Omega$, within $32^\circ$, is observed only in a few cases, and none of these
were at same time as the clustering in $\omega$.
From this line of evidence we concluded that probability to find simultaneous clusterings
in $\Omega$ and $\omega$ for any subgroup of 11 asteroids by the chance is less than $1$ per cent.
Thus, the statistical significance of the P/2006 VW$_{139}$ family is greater than $99$ per cent.    

\subsection{Physical and dynamical characteristics of P/2006 VW$_{139}$ family}
\label{ss:char}

To ascertain the physical characteristics of the members of the P/2006 VW$_{139}$ group, we search for available
data about their spectral types and albedos. Unfortunately, little is known about these 24 
objects. In particular, none of their spectral types have been determined to date.
Geometric $V$-band albedos, $p_{V}$, have been determined for four objects (Table~\ref{t:300163list}), 
all from WISE observations \citep{wise}. The average geometric albedo derived 
from these four values is $p_{V} = 0.07$.
With their low albedos and proximity to the Themis family, most
of these objects are likely C-type asteroids. The asteroid 15156 has a somewhat larger albedo 
than the other three bodies, but considering the uncertainty of this value, its albedo is still 
within the range expected for C-type asteroids. Spectroscopic observations are encouraged however
to definitively confirm the spectral types of the members of the P/2006 VW$_{139}$ family, and
characterize the family's level of physical homogeneity.

Next, we examine the dynamical characteristics of the P/2006 VW$_{139}$ group and the 
surrounding area. The positions of the most important MMRs have been already shown and discussed 
elsewhere in this paper. To better understand the role of these resonances, 
and to check whether any connection between the dynamical characteristics of the 
surrounding area and the apparent structure of the group can be established, we perform the following test:
in a box slightly larger ($3.018-3.077$~au, $0.11-0.22$, and $0.9-4.6^\circ$ 
in osculating semi-major axis, eccentricity, and inclination, respectively) than the one 
occupied by asteroids from the group, we generate 2000 randomly distributed massless test particles. The orbits
of these particles are then numerically integrated for 10~Myr, after which we calculate their synthetic
proper elements \citep{Kne2000} and Lyapunov Characteristic Exponents (LCEs).

\begin{figure}
\includegraphics[angle=-90,scale=.47]{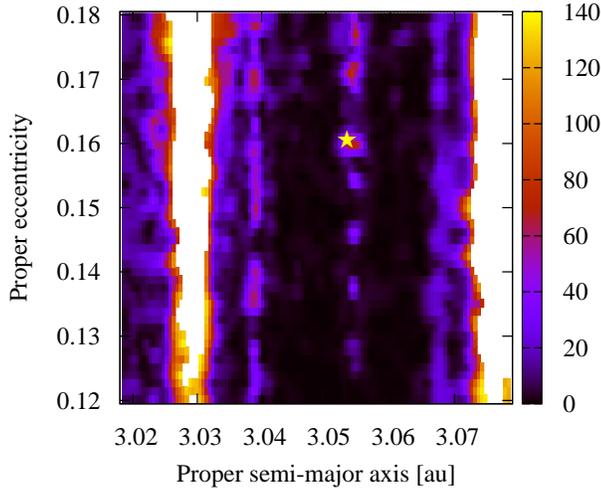}
\caption{The dynamical structure of the region around P/2006 VW$_{139}$ family.
The location of main-belt comet P/2006 VW$_{139}$ is marked by a star symbol.
The colour scale indicates Lyapunov Characteristic Exponents (LCEs) (multiplied by $10^6$) for 2,000 test particles.
The LCEs are inversely proportional to $T_{\rm lyap}$. For example, the value of 100 shown here
corresponds to $T_{\rm lyap}=10$~kyr.}
\label{f:lce_ae}
\end{figure}

\begin{figure}
\includegraphics[angle=-90,scale=.47]{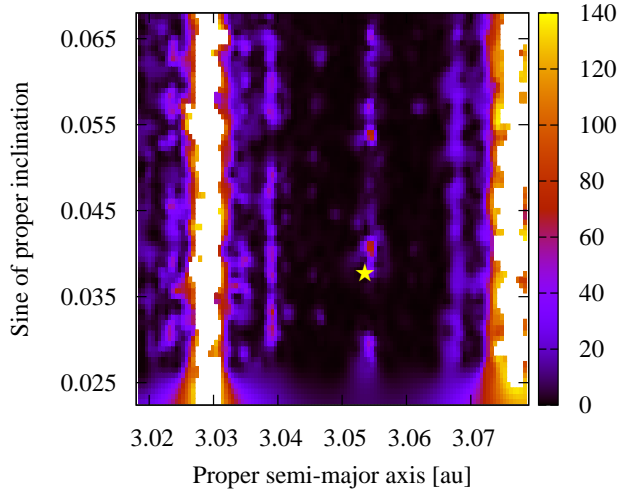}
\caption{The same as in Fig.~\ref{f:lce_ae}, but in the ($a_{p}$,$\sin(i_{p})$) plane.}
\label{f:lce_ai}
\end{figure}

\begin{figure}
\includegraphics[angle=-90,scale=.33]{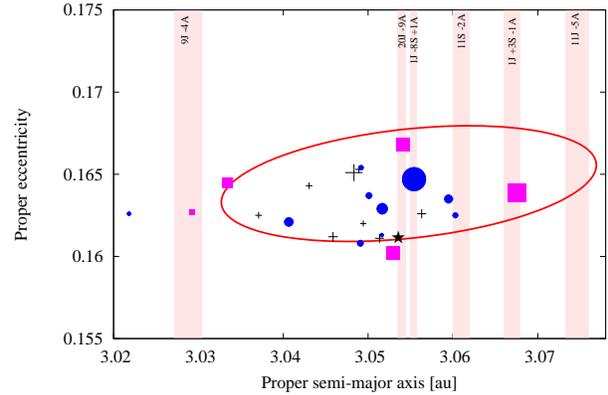}
\caption{Members of the P/2006 VW$_{139}$ group in the
($a_{p}$,$e_{p}$) plane. The superimposed ellipse marks an equivelocity curve, computed according to the
equations of Gauss \citep[e.g.][]{morby95}, for the velocity change of
$\Delta v=60$~m~s$^{-1}$, true anomaly $f=90^\circ$ 
and argument of perihelion $\omega=90^\circ$.  Circular symbols represent group members for
which $\Omega$ and $\omega$ cluster about 7.5~Myr in the past, while squares represent unstable
objects. Symbol sizes are proportional to the estimated diameters of the bodies plotted.
Vertical shaded areas mark the approximate positions of local MMRs.
The location of main-belt comet P/2006 VW$_{139}$ is denoted by the star symbol.}
\label{f:ae2}
\end{figure}

\begin{figure}
\includegraphics[angle=-90,scale=.33]{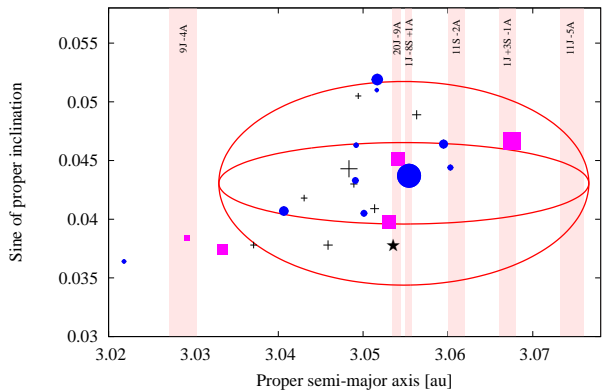}
\caption{The same as in Fig.~\ref{f:ae2} but in the ($a_{p}$,$\sin(i_{p})$) plane.
The inner ellipse corresponds to an isotropic velocity field, while the outer one is
produce assuming vertical component is 2.5 times larger than other two.}
\label{f:ai2}
\end{figure}

The results are shown in Figs.~\ref{f:lce_ae} and \ref{f:lce_ai}, where we show colour coded
maps of the obtained LCEs as functions of particle position in proper element
space. We note that the figures indicate that most of the region surrounding P/2006 VW$_{139}$ 
in orbital element space is stable,
apart from two strongly chaotic zones centred at about 3.029 and 3.075~au, corresponding to
the 9J:4A and 11J:5A MMRs, respectively. Between these two zones, there are also some smaller
areas that are weakly unstable. These weak instabilities are caused by higher
order MMRs, such as the ones shown in Figs.~\ref{f:ae2} and \ref{f:ai2}. Somewhat larger
values of the LCEs are also found in the immediate vicinity of P/2006 VW$_{139}$.  These
appear to be caused by the 20J:9A and 1J-8S+1A MMRs. However, these resonances are weak and
effective only in very narrow ranges of semi-major axis. An illustrative example is
P/2006 VW$_{139}$ itself, which is located very close to the 20J:9A resonance, but still
does not show any sign of chaotic instability. Thus, we conclude that P/2006 VW$_{139}$ family
members are largely stable and unlikely to be significantly dynamically evolved. This conclusion is also 
valid for the single family member located on the opposite side of the 9J:4A resonance from the
other family members, asteroid 2008 FG$_{14}$.

\begin{figure}
\includegraphics[angle=-90,scale=.33]{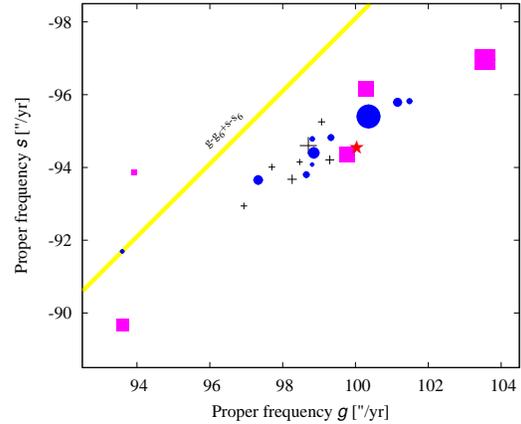}
\caption{The objects dynamically associated with the P/2006 VW$_{139}$ group plotted in the plane
of proper frequency $g$ of the longitude of perihelion \emph{versus}
proper frequency $s$ of longitude of the ascending node. As in Figs.~\ref{f:ae2} and \ref{f:ai2} circles 
represent group members whose secular angles ($\Omega$ and $\omega$) cluster, while squares represent unstable
objects. The location of main belt comet P/2006 VW$_{139}$ is marked with the star symbol.
Sizes of symbols are proportional to the estimated diameters of the corresponding objects. The inclined yellow
line indicates the position
of the $z_{1}=g-g_{6}+s-s_{6}$ secular resonance. The frequencies shown here are calculated according
to the synthetic theory \citep{Kne2000}.}
\label{f:sec_res}
\end{figure}

To develop as complete a picture of the dynamical environment as possible, we also search for the presence 
of secular resonances (up to degree six) in the region. Close to the group, we find only a non-linear
secular resonance $z_{1}=g-g_{6}+s-s_{6}$ \citep{MK1992,MilKne1994}. 
Its location is shown in Fig.~\ref{f:sec_res}. Most of the objects from the group are far enough to avoid 
interactions with this resonance. This seems to be also true for asteroid 2002 VU$_{137}$, the 
only group member located on the other side of the $g-g_{6}+s-s_{6}$ resonance 
(Fig.~\ref{f:sec_res}). However, due to the chaoticity of its orbit caused by the 9J:4A
resonance, it is difficult to estimate its secular frequencies, $g$ and $s$. Therefore, we
are unable to draw firm conclusions as to the level of influence of secular resonances on this object.
Finally, one family member, asteroid 2008 FG$_{14}$, appears to be inside the $g-g_{6}+s-s_{6}$
secular resonance.

\begin{figure} 
\includegraphics[angle=-90,scale=.32]{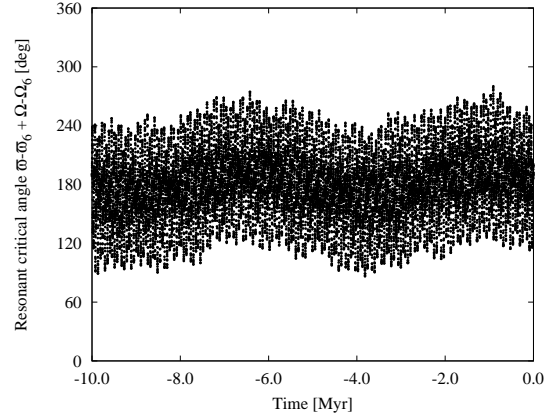}
\caption{Time evolution of the resonant critical angle $\sigma = \varpi-\varpi_{6}+\Omega-\Omega_{6}$ of the
$z_{1}=g-g_{6}+s-s_{6}$ secular resonance for asteroid 2008 FG$_{14}$.
The critical angle is in an anti-aligned libration state around $180^\circ$
over the complete time period of 10~Myr covered by the integrations.}
\label{f:crit_angle}
\end{figure}

To investigate this possibility, we plot the corresponding resonant critical angle $\sigma = \varpi-\varpi_{6}+\Omega-\Omega_{6}$ 
in Fig.~\ref{f:crit_angle}. The behaviour of $\sigma$, which oscillates around $180^\circ$, 
unequivocally confirms that 2008 FG$_{14}$ is locked within the $z_{1}$ secular resonance.
Because of that, its association to the P/2006 VW$_{139}$ family is less reliable,
meaning that it could be an interloper.

In terms of proper semi-major axis, 2008 FG$_{14}$ is located 
significantly outside the assumed equivelocity curves shown in Figs.~\ref{f:ae2} and \ref{f:ai2}.
The usual explanation for this would be Yarkovsky induced drift. However, 
the P/2006 VW$_{139}$ family seems to be young, and the role of thermal effects is not so obvious. 
The maximum possible drift speed in proper semi-major axis 
due to the Yarkovsky effect acting on a 1~km body of $(da/dt)_{max}=4 \times 10^{-4}$~au/Myr 
(see Section~\ref{ss:bim}) translates into a total drift of $3 \times 10^{-3}$~au over a period of 7.5~Myr.
As Yarkovsky induced drift is inversely proportional to the diameter of the body, 
for 2008 FG$_{14}$, it reduces to the maximum value of $1.8 \times 10^{-3}$~au 
since the family forming event. This value is almost one order of magnitude smaller
than necessary to explain the observed displacement in $a_{p}$. The obvious conclusion is 
that the Yarkovsky effect is not responsible for this object's anomalous displacement in $a_p$,
and that 2008 FG$_{14}$ could be an interloper in the P/2006 VW$_{139}$ family.

Insights into collisional physics may be obtained by analysing the inferred ejection velocity field (EVF) 
of the fragments \citep[e.g.,][]{cellino1999}. An isotropic EVF can be computed using the well known Gaussian equations.
These equations define ellipses in the ($a_{p}$,$e_{p}$) and ($a_{p}$,$\sin(i_{p})$) planes, 
whose shapes and orientations are controlled by two angles, the true anomaly ($f$) 
and argument of perihelion ($\omega$) at the epoch of break-up. To connect these two ellipses with real family members, 
it is necessary to estimate the position of the family's barycentre (i.e., its centre of mass).
Then, by adjusting the values of $f$ and $\omega$, and experimenting with different ejection 
velocity changes ($\Delta v$), it is possible to find values of each of these parameters that best correspond
to a given family, permitting us to infer some of the properties of the EVF. However, it is
well known that asteroid families evolve over time with respect to their original
post-impact configurations. They spread and become more dispersed due to 
chaotic diffusion and gravitational and non-gravitational perturbations
\citep{flora2002,carruba03,delloro04,Novakovic2010b}. Young families like P/2006 VW$_{139}$, however,
may not have had enough time since their formation to be significantly affected by these effects, meaning
that their ejection velocity fields (EVFs) may possibly be reconstructible.

The first step in attempting to reconstruct the P/2006 VW$_{139}$ family's EVF
is to estimate a position of the barycentre of the family. 
In order to do this, the absolute magnitudes ($H$; Table~\ref{t:300163list}) of the members
of the family are converted into diameters, $D$, using the standard relation \citep[e.g.][]{bowell1989}
\[
 D~{\rm (km)} = 1329~ \frac{10^{\frac{-H}{5}}}{\sqrt{p_{v}}}.
\]
Measured albedos are used for the four objects for which these are available, while the average
value of the measured albedos ($p_{v}=0.07$) is assumed for all other bodies. Then, assuming 
the same density for all objects, we estimated their masses, which are then used to derive the position 
of the barycentre in proper orbital element space. Only the 11 objects classified as real family members are used in this calculation. 
The obtained barycentre position is at $a_{p}=3.05480$~au, $e_{p}=0.1644$ and $\sin(i_{p})=0.0431$.
The fact that asteroid 2008 FG$_{14}$ may be an interloper has negligible influence on this
result.

As asteroid 15156 is the largest member of the family by far, the barycentre position is controlled by its position. 
As a consequence, the ellipses shown in Figs.\ref{f:ae2} and \ref{f:ai2} are shifted towards 
larger values of $a_p$. This may be a result of a notably asymmetric
ejection velocity field in terms of semi-major axis, caused by a large transverse velocity component. 
Such a situation often arises from cratering events \citep{Marzari1996,Novakovic2010a,VokNes2011},
though is sometimes observed in the cases of catastrophic fragmentation events as 
well \citep{Zappala1984}, although these situations are less likely.
To characterise the event which produced the P/2006 VW$_{139}$ family, we estimate the size of the parent 
body by simply summing up the volumes of all 11 members\footnote{There are 
other, generally better, methods to estimate the size of a parent body \citep{tanga99,durda2007}, but we choose
this approach here for simplicity.}. These volumes
are obtained using the diameters listed in Table~\ref{t:300163list}. Using this method, we calculate a nominal diameter
for the parent body of $D_{PB}=11.1$~km. From this value we estimate the largest remnant to parent
body mass ratio $M_{LR}/M_{PB}$ to be 0.83. 
A realistic value of $M_{LR}/M_{PB}$ is likely somewhat
smaller than the value mentioned here because the diameter of the parent body estimated from known family 
members is a lower limit \citep{durda2007}. From this evidence, we can therefore conclude
that the P/2006 VW$_{139}$ family was not formed in a completely catastrophic fragmentation event. At the moment though, it is not
possible to distinguish whether the formation of the family should be classified as a cratering or partially catastrophic event. 
In any case, the ratio $M_{LR}/M_{PB}$ is high, and may be
an explanation for observed asymmetry in the EVF. 
 
Alternatively, the calculated position of the barycentre may be wrong.
It may be affected by exclusion of unidentified
family members from our calculations. While the chance that any of eight stable asteroids which do not exhibit 
clustering belong to the P/2006 VW$_{139}$ family is quite small, the status of the five
unstable objects remains unclear.
Inclusion of these objects in the calculation of the barycentre position
moves it to $a_{p}=3.05604$~au, $e_{p}=0.1642$ and $\sin(i_{p})=0.0437$.
In this case, the ejection velocity field becomes even more asymmetric than before. The inclusion
of these 5 objects would additionally decrease the $M_{LR}/M_{PB}$ ratio. This would make
the explanation that the asymmetry of the EVF is a result of the impact event even less likely. Thus,
we believe that at least some of the unstable objects associated to the group by the 
HCM are not real family members, but we are unable to draw a firm conclusion at this time. 

Finally, despite being currently fairly stable, asteroid
15156 could have experienced some small displacement due to chaotic diffusion
as it is very close to the slightly chaotic area centred at 3.055~au (Fig.~\ref{f:lce_ae}).
Thus, it may have been less stable in the past, but then evolved onto its current stable orbit
due to chaotic diffusion and/or the Yarkovsky effect. Even an exceptionally small amount of Yarkovsky 
induced drift of the semi-major axis would be enough to drive this asteroid from an
unstable to a stable region (or vice versa). The maximum value of $(da)_{15156}$ is about $3 \times 10^{-3}$~au 
for 7.5~Myr. Hence, the position of the barycentre might not be fully preserved.

The last characteristic of the EVF that we want to draw attention to is a notably larger 
spread in the inclination than in the eccentricity (roughly by about a factor of $2$). 
This could not be related to the position of the barycentre. Of course, it could be easily
explained by a large vertical velocity component of the EVF (see the two ellipses shown in Fig.~\ref{f:ai2}).
Such situations are reported earlier \citep[e.g.][]{Zappala1984,Tsiganis2007}, but to better understand 
the possible reasons for this, identification of additional family members is necessary.
Hence, we postpone this analysis for a future work.

\section{Summary and Conclusions}

If the cluster of asteroids surrounding P/2006 VW$_{139}$ does in fact 
have a common physical origin, this object would join 133P as the second 
MBC to be associated with a young collisional family, supporting the 
hypothesis by \citet{Hsieh2009} that such associations may be necessary for 
present-day sublimation on main-belt objects to occur.  The lack of 
confirmed young family associations for any other MBCs does not 
necessarily undermine this hypothesis, as it is possible that many young 
families remain unidentified due to insufficient numbers of their members 
simply not yet being discovered, or having insufficiently well-established 
orbits to perform dynamical linkage analyses.  As more asteroids are 
discovered and orbits of known asteroids are improved by current and 
next-generation all-sky surveys (e.g. Pan-STARRS), it will be important to 
continue to search for small clusters of dynamically related asteroids, 
particularly around known MBCs, that may be previously unidentified young 
families.

Meanwhile, now that the results of our application of the BIM indicate 
that the P/2006 VW$_{139}$ group is likely a real asteroid family, it will 
be beneficial to provide observational confirmation of this dynamical 
result.  Very little information about the spectral characteristics of 
asteroids belonging to this newly identified family is currently known. 
As such, we encourage physical characterizations of the members of this 
family, either photometrically via broadband colours or spectroscopically, 
to help assess the plausibility of them being physically related in 
addition to be dynamically related.  Furthermore, since family members are 
expected to share physical properties, the fact that P/2006 VW$_{139}$ has 
been observed to exhibit cometary activity apparently due to the 
sublimation of volatile ice \citep{Hsieh2012b}, optical surveys for 
similar activity among its fellow family members may also be useful in the 
ongoing search for more MBCs.

\section*{Acknowledgments}

The work of B.N. has been supported by the Ministry of Education and Science of Serbia, under the Project 176011. 
B.N. also acknowledges support by the European Science Foundation through GREAT Exchange Grant No. 3535.
H.H.H. is supported by NASA through Hubble Fellowship grant HF-51274.01 awarded by 
the Space Telescope Science Institute, which is operated by the Association of 
Universities for Research in Astronomy (AURA) for NASA, under contract NAS 5-26555.

\bsp

\label{lastpage}

\end{document}